\documentclass{iopart}
\usepackage{bm,graphics,graphicx,epsfig,color,subfigure,rotating,latexsym,textcomp}
\usepackage{amsthm}
\usepackage{amssymb}
\usepackage{url}
\usepackage{fancyhdr}
\usepackage{alltt}
\usepackage{acronym}

\def\be{\begin{equation}}
\def\ee{\end{equation}}

\def\bea{\begin{eqnarray}}
\def\eea{\end{eqnarray}}

\def\cross{\times}

\begin{document}

\title[A coherent triggered search for single spin CBCs]{A coherent triggered search for single spin compact binary coalescences in gravitational wave data}

\author{I.\ W.\ Harry, S.\ Fairhurst}
\address{School of Physics and Astronomy, Cardiff University, Queens Buildings, The Parade, Cardiff, CF24 3AA, UK}
\eads{\mailto{ian.harry@astro.cf.ac.uk}
\mailto{Stephen.Fairhurst@astro.cf.ac.uk}}

\begin{abstract}
In this paper we present a method for conducting a coherent search for
single spin compact binary coalescences in gravitational wave data and
compare this search to the existing coincidence method for single spin
searches.  We propose a method to characterize the regions of the
parameter space where the single spin search, both coincident and
coherent, will increase detection efficiency over the existing
non-precessing search. We also show example results of the coherent
search on a stretch of data from LIGO's fourth science run but note that
a set of signal based vetoes will be needed before this search can be
run to try to make detections.
\end{abstract}


\acrodef{LIGO}[LIGO]{Laser Interferometer Gravitational-wave Observatory}
\acrodef{CBC}[CBC]{compact binary coalescence}
\acrodef{S6}[S6]{LIGO's sixth science run}
\acrodef{VSR23}[VSR2 and VSR3]{Virgo's second and third science runs}
\acrodef{EM}[EM]{electromagnetic}
\acrodef{GW}[GW]{gravitational wave}
\acrodef{NS}[NS]{neutron star}
\acrodef{BNS}[BNS]{binary neutron star}
\acrodef{NSBH}[NSBH]{neutron star-black hole binary}
\acrodef{GRB}[GRB]{gamma-ray burst}
\acrodef{S5}[S5]{LIGO's fifth science run}
\acrodef{S4}[S4]{LIGO's fourth science run}
\acrodef{VSR1}[VSR1]{Virgo's first science run}

\acrodef{PSD}[PSD]{power spectral density}
\acrodef{VSR3}[VSR3]{Virgo's third science run}
\acrodef{BBH}[BBH]{binary black holes}
\acrodef{SNR}[SNR]{signal-to-noise ratio}
\acrodef{SPA}[SPA]{stationary-phase approximation}
\acrodef{LHO}[LHO]{LIGO Hanford Observatory}
\acrodef{LLO}[LLO]{LIGO Livingston Observatory}
\acrodef{LSC}[LSC]{LIGO Scientific Collaboration}
\acrodef{PN}[PN]{Post-Newtonian}
\acrodef{DQ}[DQ]{data quality}
\acrodef{IFO}[IFO]{interferometer}
\acrodef{DTF}[DTF]{detection template families}
\acrodef{FAR}[FAR]{false alarm rate}
\acrodef{FAP}[FAP]{false alarm probability}
\acrodef{PTF}[PTF]{physical template family}

\section{Introduction}

The \ac{LIGO} and Virgo scientific collaborations have performed many
searches for \ac{CBC} signals in data taken by gravitational wave
interferometers \cite{Abadie:2010yba,Abbott:2009qj,Abbott:2009tt}. The
majority of these searches have utilized template waveforms where the
spins of the individual components are neglected. In some areas of the
parameter space spin can have a significant effect on the evolution of
the system, and consequently the emitted gravitational waveform
\cite{ACST94,Grandclement:2002dv}, leading to a poor match with the
non-spinning templates.  In these regions of parameter space the use of
templates incorporating spin will provide an increase in search
sensitivity.

Incorporating spin into template waveforms in a gravitational wave
search is a complex problem. In a non-spinning search for \acp{CBC} with
circular orbits, a source is described by nine physical parameters
\cite{Th300}. The majority of these do not affect the signal morphology,
but serve to change the overall amplitude, phase or coalescence time of
the signal and are easily maximized over \cite{Allen:2005fk}. Therefore,
template placement can be restricted to the two dimensional space of
component masses \cite{Bank06}.  A spinning \ac{CBC} in a circular
orbit, however, is described by 15 physical parameters \cite{Th300}. The
challenge is to formulate a method to detect any manner of spinning
system while limiting the number of templates, such that an analysis can
be run in a reasonable amount of time.  The problem is simplified if the
spins are aligned with the orbital angular momentum. In this case the
system will have no precession and is described by just two extra
parameters --- the spins of both bodies in the direction of the angular
momentum. Furthermore, these non-precessing waveforms are well described
by a single spin parameter \cite{Ajith:2009bn}, and it it therefore
feasible to search for non-precessing waveforms using a three
dimensional template bank.

At the time of writing only one search for \ac{CBC}s using spinning
templates with precession in LIGO/Virgo data has been published
\cite{Abbott:2007ai}.  This search utilized a phenomenological waveform
family designed to capture precessional effects \cite{BCV03b}, but was
later abandoned because it was not found to increase efficiency relative
to the non-spinning search \cite{Abbott:2009tt,Vandenbroeck:2009cv}.
This was due to the ability of the phenomenological templates to match
non-stationarities in the data and the lack of an effective signal
consistency test to veto them such as the $\chi^2$ test used in the non
spinning search \cite{Allen:2004gu}.

The \ac{PTF} waveforms proposed in \cite{BCV03b} and further explored in
\cite{PBCV04,BCPV04,Fazi:2009} give a different method for searching for
spinning binaries with precession.  This method uses
\textit{single-spin} precessing waveforms as templates. Making clever
use of maximization, it was shown \cite{PBCV04} that a \ac{PTF} search
could be performed with a four dimensional template bank: the two
masses, the magnitude of the spin and the angle between the spin and the
orbital plane.  This method is especially useful for detecting \ac{NSBH}
systems, where the spin of the neutron star would have a negligible
effect on the dynamics of the system \cite{PBCV04}.  A coincidence
search utilizing the \ac{PTF} waveforms has been developed
\cite{Fazi:2009}.  Data from each instrument is analysed separately and
only events observed with consistent time of arrival, mass and spin
parameters in more than one detector are retained.  While coincidence
requirements for non-spinning searches are well known
\cite{Robinson:2008}, it is less clear how to define coincidence when
the additional spin parameters are present.

In a coherent search \cite{Abbott:2009nc, Pai:2000zt, Bose:1999pj,
Bose:1999bp, cohnonspin} the data from all active detectors are combined
together before searching for interesting events in the combined data.
This circumvents the need for a coincidence test between events in
different detectors.  Furthermore, a coherent search offers an increased
detection efficiency over the coincident technique when more than two
detectors are active \cite{cohnonspin}.  The coherent technique is
especially useful when the sky position is known, such as when searching
for gravitational waves in coincidence with an electromagnetic
transient, such as a \ac{GRB} \cite{Collaboration:2009kk}.  Since
\ac{NSBH} and \ac{BNS} mergers are the preferred progenitor model
\cite{nakar:2007} for short \ac{GRB}, a coherent single-spin search is
ideally suited to this source. 

In this paper we describe the implementation of a coherent search for
single spin binaries with known sky location, using the \ac{PTF}
waveforms.  We briefly review the \ac{PTF} formalism before deriving the
coherent \ac{PTF} \ac{SNR}.  Due to the increased complexity of the
spinning waveforms, the coherent \ac{SNR} has a different distribution
than its non-spinning counterpart.  In particular, there is a greater
chance of obtaining a large value of the spinning \ac{SNR}, even in
Gaussian noise.  Thus there is a trade-off between the improved spinning
signal model and the increased false alarm rate at a fixed \ac{SNR}.  We
explore the single-spin \ac{CBC} parameter space to identify regions
where spin (and precession) effects are significant enough to make the
spinning search worthwhile.  We will also briefly discuss some
possibilities for vetoing background non-Gaussian transients in the data
when using the \ac{PTF} search and present results of this search run on
a short stretch of \ac{S4} data

The layout of this paper is as follows: In section
\ref{sec:singledetspin} we briefly review the single detector \ac{PTF}
search and investigate the distribution of the spinning \ac{SNR} in
Gaussian noise. In section \ref{sec:cohPTFspin} we introduce the
coherent \ac{PTF} search and investigate the distribution of the
coherent spinning \ac{SNR}.  In section \ref{sec:tospinornot} we
identify regions of the parameter space where the \ac{PTF} search offers
increased sensitivity over the non-precessing search. Section
\ref{sec:search_and_results} briefly describes our search pipeline and the
results of these methods applied to a stretch of data from \ac{S4}.

\section{Spinning Search Using Physical Template Family Waveforms}
\label{sec:singledetspin}

In this section we give a brief recap of the PTF search and its
implementation.  We also explore the expected distribution of the
spinning \ac{SNR} in Gaussian noise and compare this to that of the
non-precessing search. For a more detailed description of the PTF search
and terminology we refer the reader to \cite{PBCV04,Fazi:2009}.  We will
follow the conventions of these earlier publications as much as
possible. We will also assume that the reader is familiar with
matched-filtering techniques and its application to gravitational wave
data analysis, if not we refer the reader to
\cite{Wainstein,Allen:2005fk}.

\subsection{Single detector analysis}
\label{sec:ptfdetstat}

The likelihood ratio of there being a signal $h$ present in the data $s$
for a single detector is given by
\be \Lambda (h) = \frac{P(s|h)}{P(s|0)} \, .\ee
Assuming the noise is Gaussian, the log likelihood can be written as
\be \log \Lambda = (h|s) - \frac{1}{2}(h|h). \ee
Where we have defined the single detector inner product
\begin{equation}\label{eq:inner_product}
  (a | b) = 4 \, \mathrm{Re} \int_{0}^{\infty}
  \, \frac{\tilde{a}(f) [\tilde{b}(f)]^{\star}}{ S_{h}(f) } \,,
\end{equation}
and $S_{h}(f)$ is the \ac{PSD} of the detector. From this starting
point, $h$ must be re-expressed in such a way that it is possible to
maximize over the majority of the parameters, leaving us with only a
small number of dimensions over which to carry out a templated search.

The dominant harmonic of the gravitational waveform can be expanded in
terms of the five $l=2$ spin-weighted spherical harmonics.  The
amplitude of each of these terms will depend upon the distance to the
source, $D$; the sky location of the source, $(\theta, \psi)$; the
orientation of the source, which is described by three angles: the
inclination, $\iota$, polarization, $\phi$, and the orientation of the
spin in the orbital plane, $\varphi$. The waveform for each of these
harmonics depends on the two masses, ($M_1,M_2$); the amplitude of the
spin, $\chi$; the angle between the spin and the orbital plane,
$\kappa$; the inital orbital phase relative to the spin direction,
$\Phi_0$, and the time of coalescence, $t_c$. Consequently 
the gravitational waveform for a single spin binary can be expressed as
\be 
h(t) = \sum_{I=1}^5 P_I(D,\theta,\phi,\psi,\iota,\varphi) 
Q^I(M_1,M_2,\chi,\kappa,\Phi_0,t_c). 
\ee
where $P_I$ are five amplitudes and $Q^I$ describe five waveform components.

To obtain the PTF detection statistic, a free maximization is carried
out over the five $P_I$ components as well as the initial orbital phase of
the system.  The \ac{SNR} will then depend on 10 components:
the 0 and $\frac{\pi}{2}$ phases of the five $Q^I$ waveforms.
Specifically, we calculate the inner product between each component
$Q^{I}$
of the waveform and the data
\begin{equation}\label{eq:ab_def}
  A^I = (s|Q^I_0) \quad \mbox{and} \quad B^I = (s|Q^I_{\frac{\pi}{2}}), 
\end{equation}
as well as between the different $Q^{I}$ themselves%
\footnote{We have made the standard assumption that $Q^I_0 = i
Q^I_{\frac{\pi}{2}}$.}
\be M^{IJ} = (Q_0^I | Q_0^J) = (Q_{\frac{\pi}{2}}^I |
Q_{\frac{\pi}{2}}^J) \, . \ee
The maximized PTF detection statistic is given by \cite{PBCV04,Fazi:2009}
\bea \label{eq:singlestat}
\frac{\rho^2}{2} = \log \Lambda |_{\rm max} &=& \frac{1}{2} \biggl[\mathbf{A}^T\mathbf{M}^{-1}\mathbf{A} + \mathbf{B}^T\mathbf{M}^{-1}\mathbf{B} \biggr] \nonumber \\
&+& \frac{1}{2}\sqrt{\left(\mathbf{A}^T\mathbf{M}^{-1}\mathbf{A} - \mathbf{B}^T\mathbf{M}^{-1}\mathbf{B}\right)^2 + \left(2\mathbf{A}^T\mathbf{M}^{-1}\mathbf{B}\right)^2}. 
\eea

The expression for the \ac{SNR}, $\rho$, can be simplified by performing
a transformation such that both $Q_0^I$ and $Q_{\frac{\pi}{2}}^I$ are
orthonormal.  First, perform a rotation on the $Q_0^I$ to make $M^{IJ}$
diagonal, then normalize the basis vectors.  We denote the orthonormal
basis $\widetilde{Q}_0^I$.  This transformation will also
orthonormalize $\widetilde{Q}_{\frac{\pi}{2}}^I$ and render
$\widetilde{M}^{IJ}$ the identity matrix.  After this transformation,
the \ac{SNR} can be written as
\begin{equation} \label{eq:singlestatsimp}
\rho^2 = \left(\widetilde{A}\cdot\widetilde{A} + \widetilde{B}\cdot\widetilde{B} \right)
+ \sqrt{\left(\widetilde{A}\cdot\widetilde{A} - \widetilde{B}\cdot\widetilde{B}\right)^2 + 
\left(2\widetilde{A}\cdot\widetilde{B}\right)^2}. 
\end{equation}
where $\widetilde{A}$ and $\widetilde{B}$ are defined as in
(\ref{eq:ab_def}).

We have performed a free maximization over the five $P_{I}$ amplitudes.
In principle, these depend upon \textit{six} physical parameters.
However, these parameters only enter in four different combinations as
\begin{itemize}

\item an amplitude parameter, dependent on ($D,\theta, \psi,\phi$),

\item the relative sensitivity of the instrument to the $+$ and $\cross$
polarizations, dependent on ($\theta, \psi,\phi$)

\item the inclination angle, $\iota$

\item the spin orientation, $\varphi$.

\end{itemize}
Therefore performing a free maximization over the five $P_I$ components
means that the maximized $P_{I}$ values may not correspond to a physical
set of parameters.  This is discussed in \cite{PBCV04} and various
methods for projecting onto the physical sub-space have been proposed.
For the case of an externally triggered search, where the sky location
is known, the situation is unchanged as the $P_I$ are still described by
the same four unknown parameters.

When the orbital plane of the system does not precess, there is
gravitational wave emission in only two of the harmonics, $Q^{1}$ and
$Q^{2}$. The other components vanish identically.  Furthermore, these two
harmonics are related by a phase shift: $Q^1 = iQ^2$.  Thus, the matrix
$M$ is degenerate and the PTF maximization breaks down.  It is, however,
straightforward to maximize over the two remaining amplitudes, and
obtain the \ac{SNR} as
\begin{equation}\label{eq:snr_no_precess}
  \rho^{2} = \frac{(s|Q^{1}_{0})^{2} + (s|Q^{1}_{\frac{\pi}{2}})^{2}}{
(Q^{1}_{0} | Q^{1}_{0})}
\end{equation}
This is identical to the well known \ac{SNR} for the non-spinning search
\cite{Allen:2005fk}, and the two phases of $Q^1$ correspond to the 0 and
$\pi/2$ phases of the non-precessing template.

The \ac{PTF} search allows one to perform a search using single spin
waveforms in a reasonable amount of time on a single detector
\cite{Fazi:2009}.  Any event with an \ac{SNR} above some preset
threshold constitutes a single detector ``trigger'', and candidate
events would be required to be observed in more than one detector.
However, it remains a challenge to derive a metric on the four
dimensional mass and spin space that could be used in generating a
template bank and in defining coincidence requirements. Furthermore, a
strategy for vetoing non-transient glitches has been suggested
\cite{Fazi:2009}, such a strategy would be needed to make a coincident
\ac{PTF} search viable.

\subsection{SNR distribution in Gaussian noise}
\label{sec:detstatdist}

The \ac{PTF} template waveform will provide a better match than a
non-spinning template to a gravitational wave signal from a spinning
binary.  However, we pay a price since we must filter the data against
more waveform components, $Q^{I}$, thereby increasing the chance of a
spurious match with the noise.  Additionally, the spinning \ac{SNR}
takes a more complex form (\ref{eq:singlestatsimp}) than the simple
quadratic expression (\ref{eq:snr_no_precess}) when there is no
precession.  Here, we will investigate the \ac{SNR} distributions in
Gaussian noise for these two cases.  In section \ref{sec:tospinornot} we
use this to identify regions of parameter space with sufficient spin
effects to warrant the use of the \ac{PTF} search.

Ten filters are used in the calculation of the PTF detection statistic:
$(\widetilde{Q}_0^I,s)$ and $(\widetilde{Q}_{\frac{\pi}{2}}^I,s)$. As
both $\widetilde{Q}_0^I$ and $\widetilde{Q}_{\frac{\pi}{2}}^I$ are
orthonormal, the only remaining freedom is the relation between the
$\widetilde{Q}_0^I$ and $\widetilde{Q}_{\frac{\pi}{2}}^I$ terms, 
\be 
\widetilde{N}^{IJ} = (\widetilde{Q}_0^I,\widetilde{Q}_{\frac{\pi}{2}}^J). 
\ee
This $\widetilde{N}^{IJ}$ is a will be a $5\times5$ antisymmetric matrix
which can have 4 non-zero eigenvalues: $\pm \lambda_1,\pm \lambda_2$.
The values of these eigenvalues determine the distribution of the
\ac{PTF} detection statistic.

\begin{figure}[htp]
  \hspace{0.25\linewidth}
  \begin{minipage}[t]{0.5\linewidth}
    \includegraphics[width=\linewidth]{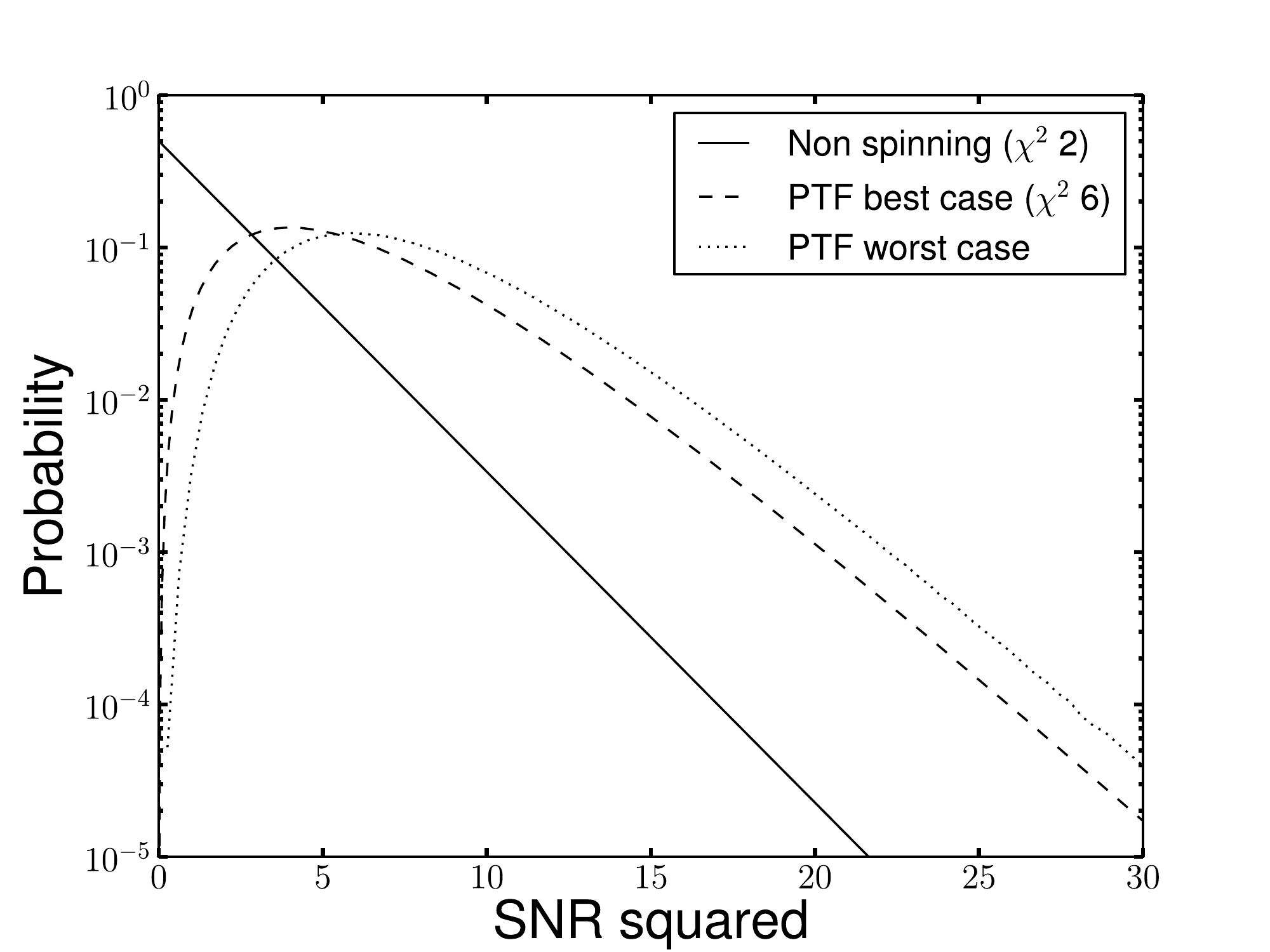}
  \end{minipage}
  \caption{\label{fig:snr_dist1} 
The ``best'' and ``worst'' possible distributions of
the single detector \ac{PTF} \ac{SNR} squared, this is compared with the non
precessing \ac{SNR} squared.  }
\end{figure}

For every \ac{NSBH} waveform we have tested using the initial LIGO
sensitivity curve, the magnitudes of $\lambda_{1}$ and $\lambda_{2}$
have been very close to 1.  Thus, although there are ten different
waveform components, we find that, in effect, only six of these are
independent --- the others are linear combinations of these six.  There
are then only six independent filters and it is not difficult to show
that the spinning \ac{SNR} (\ref{eq:singlestatsimp}) collapses to a
quadratic form which is $\chi^2$ distributed with six degrees of freedom
in Gaussian noise. This is the ``best'' case for the detection
statistic.  The ``worst'' case occurs when both $\lambda_{1}$ and
$\lambda_{2}$ are zero and all ten of the filters are independent.  In
this case, the \ac{SNR} expression cannot be simplified and its
distribution does \textit{not} correspond to a $\chi^2$ distribution
with 10 degrees of freedom as might be expected; the real distribution
is somewhat more complex. Both best and worst cases are illustrated in
Figure \ref{fig:snr_dist1}.

The \ac{SNR} (\ref{eq:snr_no_precess}) for a non-precessing template
follows a $\chi^{2}$ distribution with two degrees of freedom in
Gaussian noise.  This is also plotted on Figure \ref{fig:snr_dist1}.  By
comparing the distributions of the \ac{PTF} and non-precessing
\ac{SNR}s, it is clear that the background triggers produced by the PTF
search will have, on the average, considerably larger SNR than those
produced by the non-precessing search. We explore the effect that this
has on a search further in section \ref{sec:tospinornot}.

\section{Coherent Spinning PTF search}
\label{sec:cohPTFspin}

In this section, we introduce a multi-detector, coherent formulation of
the \ac{PTF} search.  As in \cite{cohnonspin}, we will restrict
attention to the case where the sky location is known. This simplifies a
coherent search as the sensitivity of the detectors to the two GW
polarizations and the relative time delays between detectors are known.
Astrophysically, this is of interest when searching for gravitational
waves associated to electromagnetic transients such as \acp{GRB}
\cite{nakar:2007, Shibata:2007zm}.

The data from various detectors are combined together coherently to form
two coherent data streams, with one stream containing the $+$
polarization of any gravitational wave signal present in the network and
the second containing the $\times$ polarization.  The coherent method
will offer an improvement in sensitivity over the coincidence method
when more than two detectors are used, as only two data streams are
searched.  For networks with greater than two detectors, it is also
possible to construct null streams which will contain no gravitational
wave signal, and can be used as a consistency test \cite{Guersel:1989th,
cohnonspin}.

We begin by formulating the coherent \ac{SNR} for the spinning \ac{PTF}
search and go on to
explore how this will be distributed in Gaussian noise.

\subsection{Coherent PTF Search Method}
\label{sec:spindetstat}

To formulate a coherent detection statistic for the PTF templates we
draw on many of the methods and techniques that were used in deriving
the single detector statistic described in section \ref{sec:ptfdetstat}
and derived in detail in \cite{PBCV04,Fazi:2009}. We follow the
conventions of \cite{cohnonspin} in extending this to a coherent,
multi-detector search.  Assuming that the noise in different detectors
is independent, the multi-detector likelihood is given by
\be 
\ln \Lambda =  (\mathbf{h}|\mathbf{s}) - \frac{1}{2} (\mathbf{h}|\mathbf{h}) 
\ee
where we have defined a multiple detector matched filter
\be (\mathbf{a}|\mathbf{b}) = \sum_X (a^X | b^X). \ee 
and the index $X$ runs over the detectors in the network.
 
As before, we want to maximize over as many of the parameters as
possible to minimize the dimension of the required template bank.  We
start by maximizing this over the distance, D, and initial orbital
phase, $\Phi_0$, to obtain%
\footnote{The $Y$ subscript in the inner product in the denominator
denotes the fact that the \ac{PSD} of detector $Y$ is used in evaluating
the inner product.  We do not require the noise \acp{PSD} of the
different detectors to be the same}
\begin{equation}\label{eq:likelihood_dphi}
 \ln \Lambda |_{\rm{max}(D,\Phi_0)} = 
 \frac{1}{2} \frac{ \sum_{X,I} \left[ \left( P_I^X A_I^X \right)^2 + \left( P_I^X B_I^X \right)^2 \right]}{ 
 \sum_{Y,J,K} \left[P_J^Y P_K^Y (Q_0^{J} | Q_0^K)_Y \right]}, 
\end{equation}
where $P_{I}$ are the amplitudes of the various waveform components
$Q^{I}$, and $A_{X}^{I}$, $B^{I}_{X}$ are defined as in
(\ref{eq:ab_def}).  Although the $P_{I}$ depend upon $D$ the maximized
likelihood is independent of it as scaling the distance has an identical
effect on both the numerator and denominator of
eq.~(\ref{eq:likelihood_dphi}).

As in the single detector case, we would like to maximize over the
$P_{I}$ to eliminate them.  However, in the multi-detector case, they
are detector dependent since the sensitivity of the detectors to the $+$
and $\times$ gravitational wave polarizations will differ.  These
sensitivities are encoded in the detector response functions, $F_+$ and
$F_{\cross}$, which depend on the sky location of the source in the
detector frame.  As we are focusing on an externally triggered search,
where the sky location is known, these values will be known for each
detector.  We can then factor the detector dependent terms out of the
$P_{I}^{X}$ as
\be 
P_I^X = F_+^X(\theta,\phi) S_I(D,\iota,\psi,\varphi_0) 
  + F_{\cross}^X(\theta,\phi) T_I (D,\iota,\psi,\varphi_0) 
\ee
where $S_I$ and $T_I$ denote the amplitude of the $+$ and $\cross$
components respectively of the 5 $Q_I$ in the radiation frame.  They
depend on the distance, $D$ and the angles $(\iota,\phi,\psi_0)$ that
describe the rotations necessary to transform from the source frame to
the radiation frame.  

We can re-cast the log-likelihood into a form which more closely
resembles the single detector case by introducing ten-dimensional
analogues of the $P_{I}$ and $Q^{I}$ by defining
\bea 
\mathcal{P}_{\alpha} &:=& \left[ S_1,S_2,S_3,S_4,S_5,T_1,T_2,T_3,T_4,T_5\right] \nonumber \\
\mathcal{Q}_{0,\frac{\pi}{2}}^{\alpha} &:=& \left[F_+
Q_{0,\frac{\pi}{2}}^1;...;F_+ Q_{0,\frac{\pi}{2}}^5;F_{\cross}
Q_{0,\frac{\pi}{2}}^1;...;F_{\cross} Q_{0,\frac{\pi}{2}}^5\right] \, .
\eea
The change to ten dimensions naturally arises because a multiple detector
coherent network is sensitive to both the $+$ and $\cross$ components,
whereas a single detector network is only sensitive to one polarization.
We also define the multi-detector inner products between signal and
waveform components 
\bea
\mathcal{A}^{\alpha} &=& (\mathbf{s} |\boldsymbol{\mathcal{Q}}_0^{\alpha}) 
\quad \mbox{and} \quad
\mathcal{B}^{\alpha} = (\mathbf{s} | \boldsymbol{\mathcal{Q}}_{\frac{\pi}{2}}^{\alpha}) 
\nonumber \\
\mathcal{M}^{\alpha \beta} &=&
(\boldsymbol{\mathcal{Q}}_0^{\alpha} | \boldsymbol{\mathcal{Q}}_0^{\beta})
=
(\boldsymbol{\mathcal{Q}}_{\frac{\pi}{2}}^{\alpha} | \boldsymbol{\mathcal{Q}}_{\frac{\pi}{2}}^{\beta}) \nonumber .
\eea

The log likelihood equation can then be written as
\begin{equation}\label{eq:cohlikeone}
\ln \Lambda |_{\rm{max}(D,\Phi_0)} = \frac{1}{2} 
  \frac{\mathcal{P}_{\alpha} \mathcal{P}_{\beta} 
  \left(\mathcal{A}^{\alpha}\mathcal{A}^{\beta} + 
  \mathcal{B}^{\alpha}\mathcal{B}^{\beta}\right)}{ 
\mathcal{P}_{\alpha} \mathcal{P}_{\beta} \mathcal{M}^{\alpha \beta}}
\end{equation} 
We proceed, as before, by transforming to an orthonormal basis
$\widetilde{\mathcal{Q}}_{0}^{\alpha},
\widetilde{\mathcal{Q}}_{\frac{\pi}{2}}^{\alpha}$ for the waveform
components.  Then, maximizing freely over $\mathcal{P}_{\alpha}$ yields
the coherent \ac{PTF} \ac{SNR}
\begin{equation}\label{eq:cohsnrsq}
\rho_{\mathrm{coh}}^2 = \left[\widetilde{\mathcal{A}} \cdot \widetilde{\mathcal{A}} + \widetilde{\mathcal{B}}\cdot\widetilde{\mathcal{B}} \right] 
+ \sqrt{\left(\widetilde{\mathcal{A}}\cdot\widetilde{\mathcal{A}} - \widetilde{\mathcal{B}}\cdot\widetilde{\mathcal{B}}\right)^2 + \left(2\widetilde{\mathcal{A}}\cdot\widetilde{\mathcal{B}}\right)^2},
\end{equation}
where, as before, the tilde denotes that we are in the orthonormal
basis.  This is very similar in form to the single detector statistic in
equation (\ref{eq:singlestat}).  When the network is only sensitive to
one polarization, the matrix $\mathcal{M}^{\alpha \beta}$ becomes
degenerate and the maximization procedure must be re-visited.  Here it
is natural to remove all terms corresponding to the second 
polarization and reduce to 5 dimensions, as in the single detection
search. Additionally, in section \ref{sec:ptfdetstat} we noted that
when the template has no precession the single detection \ac{PTF} SNR collapses
to the familiar SNR formalism used in the non-spinning search. Similarly,
in the coherent \ac{PTF} search, when the template has no precession,
the coherent SNR will collapse to the non-spinning coherent SNR given
in \cite{cohnonspin}.

The coherent \ac{SNR} of equation (\ref{eq:cohsnrsq}) can be used as a
detection statistic in performing a coherent search using PTF templates,
as we explore in section \ref{sec:search_and_results}.  In the single
detector search, we maximized freely over five $P_{I}$ which were
dependent upon four physical parameters.  Here, the
$\mathcal{P}_{\alpha}$ still depend on only 4 parameters but we are now
maximizing over ten amplitudes.  This clearly introduces a lot of
unnecessary freedom.  We are currently investigating alternative methods
of constructing the coherent \ac{SNR} which might eliminate these
un-physical degrees of freedom.  However, we should note that the
coincidence search allows for a similar freedom as the $P_{I}$ are
maximized independently for each detector.  Consequently, for a network
with three or more detectors, the coherent search provides a sensitivity
improvement.

\subsection{SNR distribution in Gaussian noise}
\label{sec:cohdist}

In section \ref{sec:detstatdist} we explored how the single detector PTF
statistic is distributed in Gaussian noise.  For the coherent PTF search
we can use a similar strategy to investigate the distribution of the
coherent \ac{SNR}. In the coherent case there are twenty filters
$\widetilde{\mathcal{A}}^{\alpha}$ and
$\widetilde{\mathcal{B}}^{\alpha}$ and we have constructed the detection
statistic such that
$\widetilde{\mathcal{Q}}_{0}^{\alpha}$ and
$\widetilde{\mathcal{Q}}_{\frac{\pi}{2}}^{\alpha}$ are orthonormal. As
before, the only freedom is the relationship between the $0$ and
$\frac{\pi}{2}$ terms encoded in
\be \widetilde{\mathcal{N}}^{\alpha\beta} = (\widetilde{\boldsymbol{\mathcal{Q}}}_0^{\alpha}
,\widetilde{\boldsymbol{\mathcal{Q}}}_{\frac{\pi}{2}}^{\beta}). \ee
\begin{figure}[t]
  \hspace{0.25\linewidth}
  \begin{minipage}[t]{0.5\linewidth}
    \includegraphics[width=\linewidth]{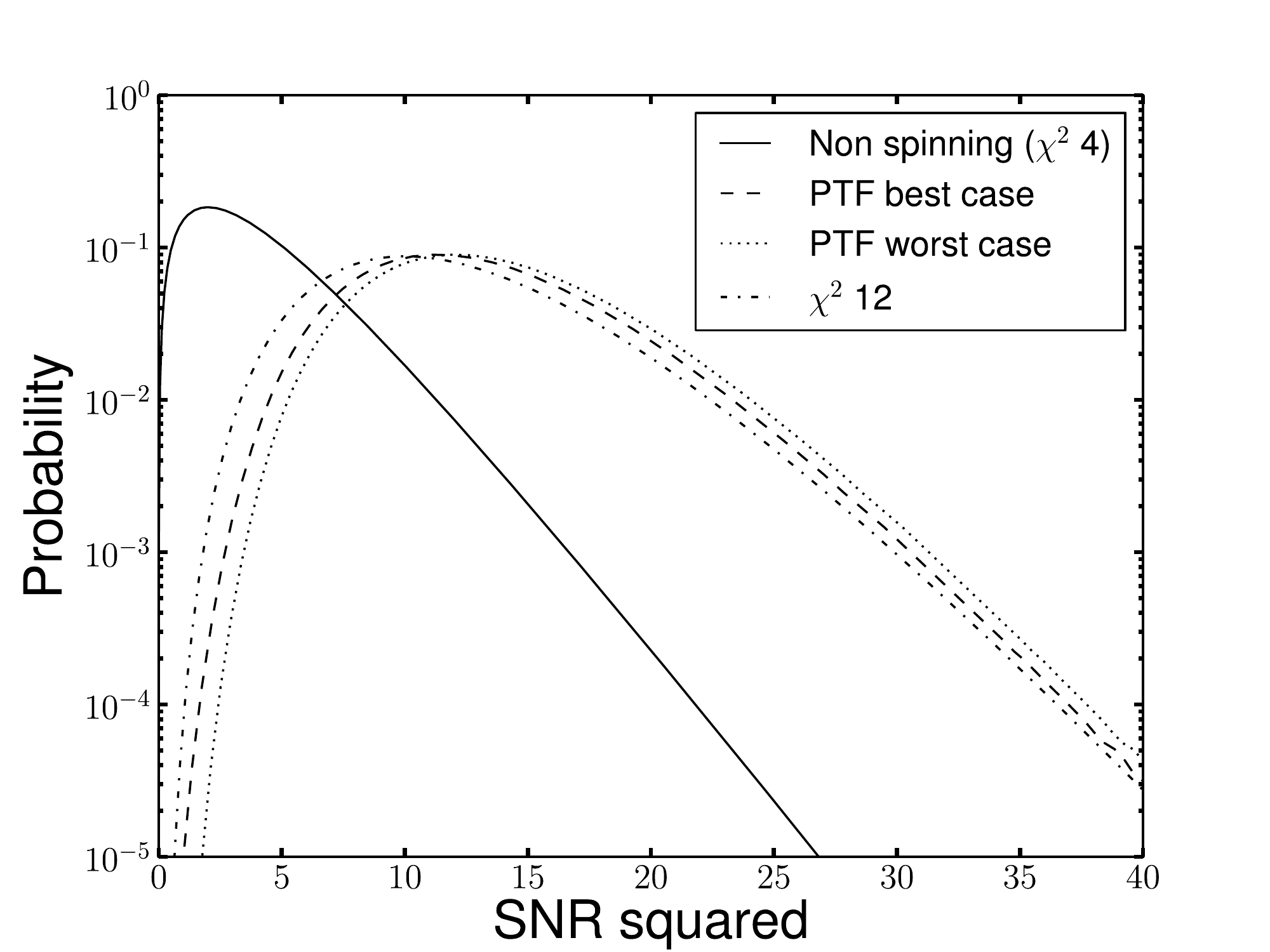}
  \end{minipage}
  \caption{\label{fig:snr_dist2} 
The ``best'' and
``worst'' possible distributions of the coherent \ac{PTF} \ac{SNR} squared 
as well as the distribution of the non spinning \ac{SNR} squared.  }
\end{figure}
This is a 10x10 antisymmetric matrix comprised of four 5x5 blocks, each
of which is antisymmetric. Therefore this matrix can have 8 non-zero
eigenvalues: $\pm \lambda_{1,2,3,4}$.  These eigenvalues determine the
distribution of coherent \ac{SNR} in Gaussian noise --- for smaller
eigenvalues, the large \ac{SNR} tail of the distribution becomes more
significant.  In the tests that we have performed using the initial LIGO
sensitivity curve and \ac{NSBH} precessing templates, all four
eigenvalues give values close to unity, the ``best'' case in which there
are 12 independent waveform components.  However, the distribution does
not collapse to a $\chi^2$. In Figure \ref{fig:snr_dist2}, we demonstrate
that this gives a distribution {\it similar} to a $\chi^2$ distribution
with 12 degrees of freedom. In the ``worst'' case, where all of the
eigenvalues are equal to 0, there are 20 independent waveform components
and this distribution is also shown in Figure \ref{fig:snr_dist2}.

\section{Identifying where the PTF search is most beneficial}
\label{sec:tospinornot}

In sections \ref{sec:singledetspin} and \ref{sec:cohPTFspin} we have
derived the spinning \ac{SNR} that can be used to perform a
gravitational wave search using single spin inspiral waveforms as
templates. We have demonstrated that, on the average, background
triggers will have larger values of SNR in the \ac{PTF} search than in
the non-precessing search. At the same time, precessing \ac{PTF}
waveforms will be a better match to any spinning, precessing signals in
the data.  This begs the question as to whether it is preferable to use
a search with non-precessing waveforms or  single spin \ac{PTF}
waveforms to detect precessing systems.  The PTF triggers will match the
waveform better but this comes at the cost of searching a larger
parameter space.

To quantify this, in Table \ref{tab:snrs} we give the \ac{SNR} that
corresponds to a \ac{FAP} of $10^{-10}$ in Guassian noise for the
various searches. We chose this value because it roughly corresponds to
the loudest background events we observe when running the search on 2000
seconds of Gaussian noise, as is appropriate for a \ac{GRB} search. The
figures in the table show that the \ac{PTF} search must obtain 26\% more
signal power (\ac{SNR} squared) to be more efficient in the single
detector case at this FAP and 38\% more signal power for the coherent
case.

\begin{table}
\begin{center}
 \begin{tabular}{c|cc}
  \hline
  \hline
   & Non Precessing & \ac{PTF} \\
  \hline
  Single detector & 6.79 & 7.63 \\
  Coherent        & 7.26 & 8.53 \\
  \hline
  \hline
 \end{tabular}
 \caption{\label{tab:snrs} 
The \ac{SNR} corresponding to a \ac{FAP} of $10^{-10}$ for the non
precessing and the \ac{PTF} search, for both coherent and single
detector cases. Here, for the \ac{PTF} case the single detector
and coherent detection statistics are assumed to be $\chi^2$
distributions with 6 and 12 degrees of freedom respectively.}
\end{center}
\end{table}

There are large areas of the parameter space where precession will not
significantly effect the evolution of the binary and thus a
non-precessing template will  pick up the majority of the power in a
precessing signal. In these areas it would be better to search for the
spinning signal with a non-precessing template, achieving a lower
\ac{FAP} than for the \ac{PTF} search using an exactly matching
template.  Equivalently, when a system has little precession, the
majority of power is contained in the $Q^{1}$ and $Q^{2}$ components of
the \ac{PTF} waveform and these two components are very similar, up to
an overall phase shift.  We can then consider performing a ``restricted
PTF'' search, where we filter only these two components of the waveform
against the data.  This serves to reduce the \ac{FAP} at a fixed
\ac{SNR} while losing only a small amount of the power in the signal.

To do this, we test every template waveform, before filtering, to
determine whether on whether the template would be more likely to detect
a matching signal below a false alarm probability of $10^{-10}$ using
the restricted or full \ac{PTF} search.  This can be calculated by
simulating a large number of gravitational wave signals, with masses and
spin matching those of the template, but uniformly distributed in volume
and orientation.  Then, simply count the number of simulated signals
expected to give an \ac{SNR} greater than the value corresponding to a
\ac{FAP} $10^{-10}$ (given in Table \ref{tab:snrs}) for both methods.
Whichever of the PTF or restricted methods is expected to perform better
is then used when filtering the data with that template.  Using this
method, we are able search the full parameter space of \ac{NSBH}
binaries in a single search, including non-spinning, non-precessing,
marginally precessing and fully precessing configurations.  This method
works equally well for the single detector or the coherent search. 

\begin{figure}[t]
  \hspace{0.25\linewidth}
  \begin{minipage}[t]{0.5\linewidth}
    \includegraphics[width=\linewidth]{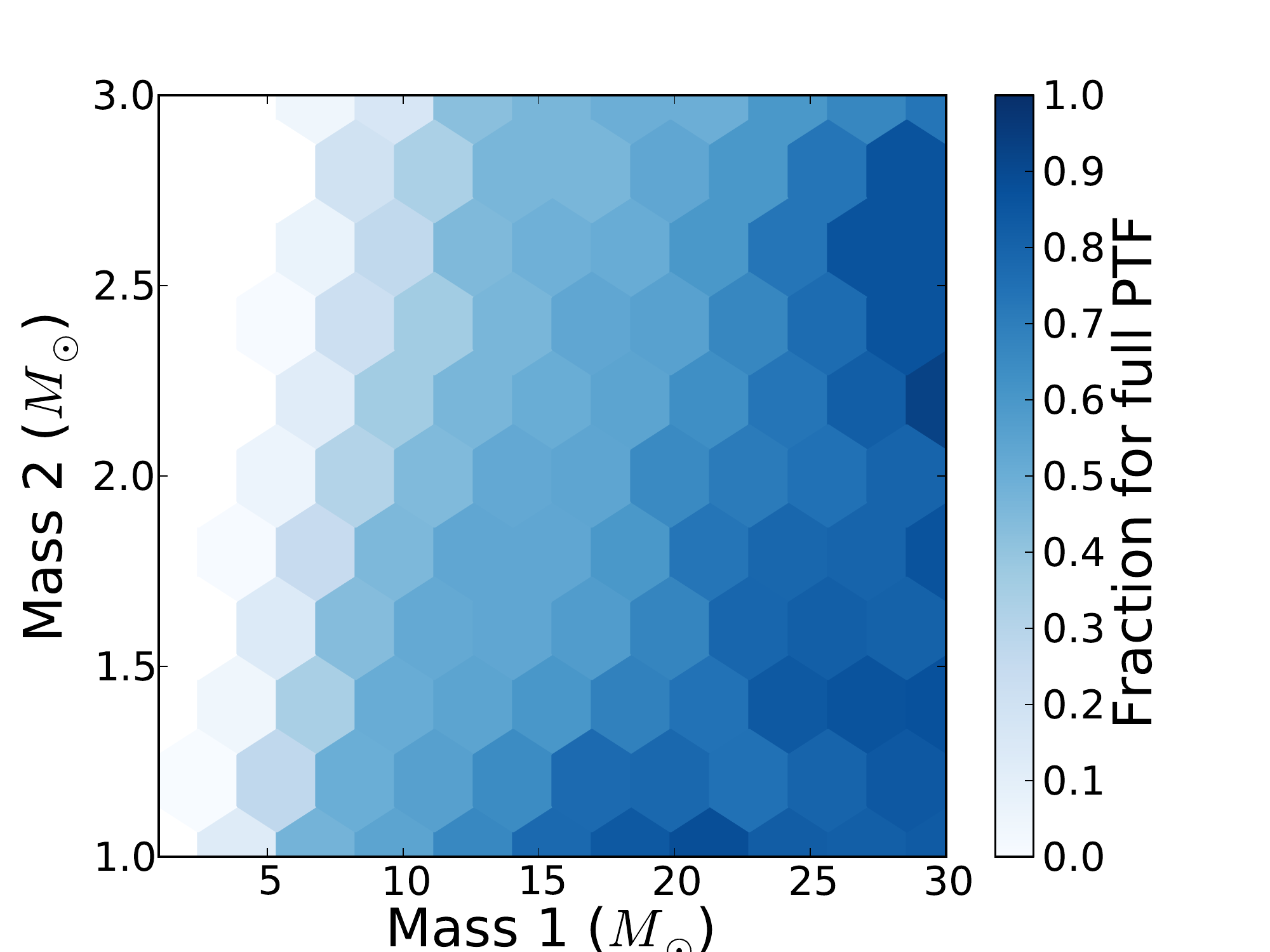}
  \end{minipage}
  \caption{\label{fig:S4banksplit} 
The fraction of templates analysed by the full PTF statistic as a
function of the masses in the \ac{NSBH} region of the parameter space.
}
\end{figure}

In Figure \ref{fig:S4banksplit} we illustrate the fraction of templates
analysed by the full PTF statistic, as a function of the masses, for the
coherent search.  The splitting of the templates into full and
restricted does not require filtering against the data, but it does make
use of the \acp{PSD} of the detectors.  For this study, we use data from
the three \ac{LIGO} detectors during the \ac{S4} run. 

A template bank was generating by taking a standard non-spinning
template bank \cite{Cokelaer:2007kx} in the mass space and, for each value of
the masses, creating 15 templates with identical masses but spin
parameters gridded over the two dimensional spin space, as described
in \cite{Fazi:2009}.  The precessing
single spin templates are most needed in the high mass ratio region of the
parameter space.  For this template bank, there are 35395 templates to
be analysed with the restricted method and 14660 templates to be
analysed with the full \ac{PTF} method.

\section{Search method and example results}
\label{sec:search_and_results}

In section \ref{sec:cohPTFspin} we derived a detection statistic
appropriate for a coherent search using the \ac{PTF}
waveforms as templates. In section \ref{sec:tospinornot} we described a method
through which one can identify where the \ac{PTF} search is most needed
and to split a template bank into those templates that should be
analysed with the full \ac{PTF} statistic and those that should be
analysed with the restricted \ac{PTF} statistic. We have combined these
two methods together to create a search pipeline than can be used to
coherently analyse gravitational wave data to search for precessing
\ac{NSBH} signals associated to short \acp{GRB}.  We will briefly
describe the analysis procedure before presenting an example result.

The search uses much of the same architecture as that described in
\cite{cohnonspin} and \cite{Abadie:2010uf}. Namely we search for
gravitational wave signals in the ``on-source'' time, defined to be
[-5,+1) seconds around the reported time of the \ac{GRB}. Background is
estimated from performing 324, 6 second trials around the \ac{GRB} time,
but at least 48s away from the on-source time.  The coherent \ac{PTF}
search makes use of the same infrastructure as a coherent
non-spinning search described in \cite{cohnonspin}.  In particular, the data
handling, \ac{PSD} estimation and matched filtering routines are the
same.  Of course, the coherent \ac{PTF} search makes use of spinning,
precessing waveforms in the filtering and computes the \ac{SNR} given in
equation (\ref{eq:cohsnrsq}).

To demonstrate the performance of the coherent \ac{PTF}, we ran it over
a stretch of data from \acp{LIGO} \ac{S4} run.  The data was chosen
randomly, subject to the condition that all three of the \ac{LIGO}
detectors were operational at the time.  This is the same data as was
used to illustrate the template bank splitting in section
\ref{sec:tospinornot}, and the same bank with 15,000 full \ac{PTF} and
35,000 restricted \ac{PTF} templates was used. 

\begin{figure}[t]
  \begin{minipage}[t]{0.495\linewidth}
    \includegraphics[width=\linewidth]{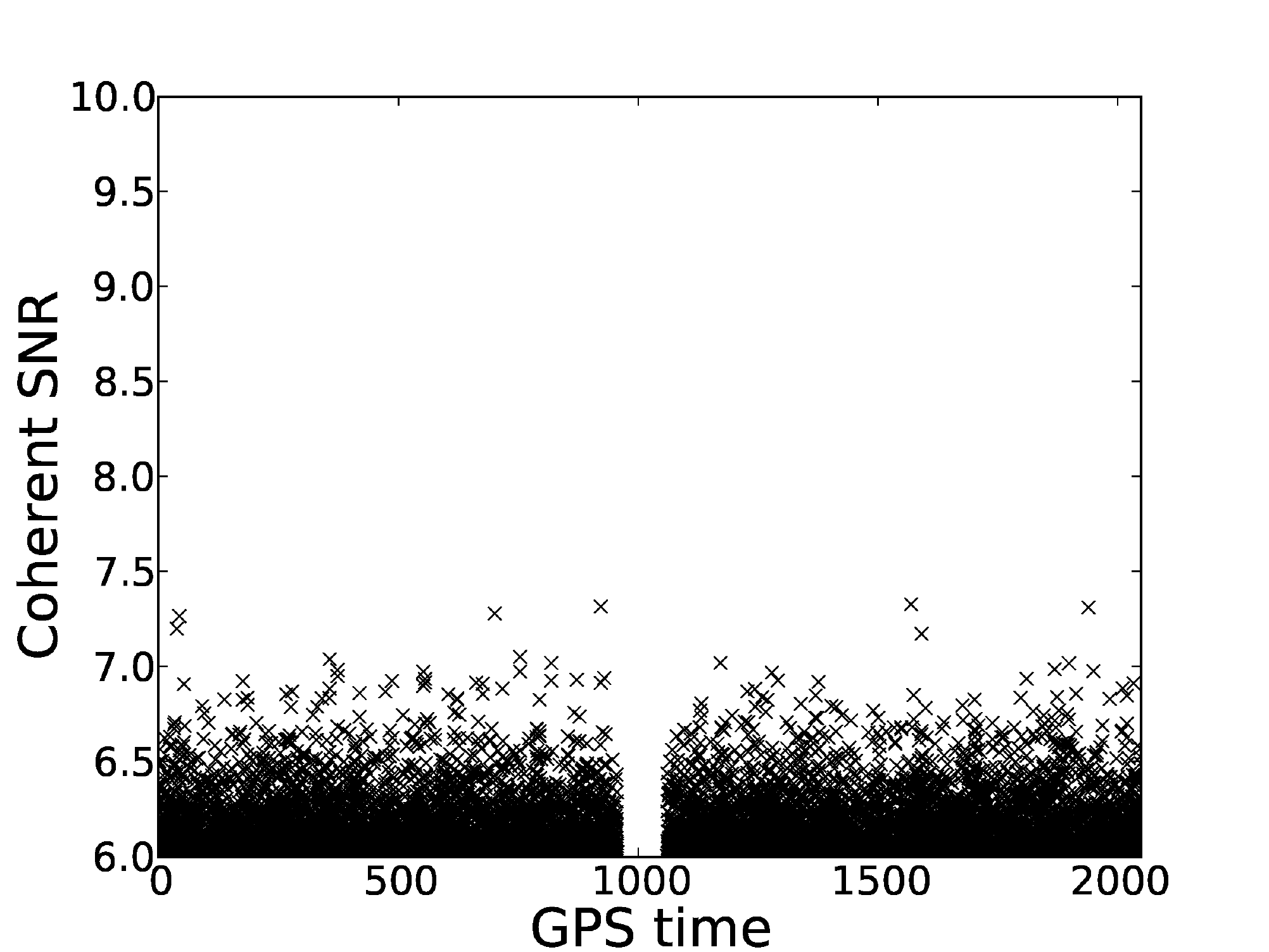}
  \end{minipage}
  \begin{minipage}[t]{0.495\linewidth}
    \includegraphics[width=\linewidth]{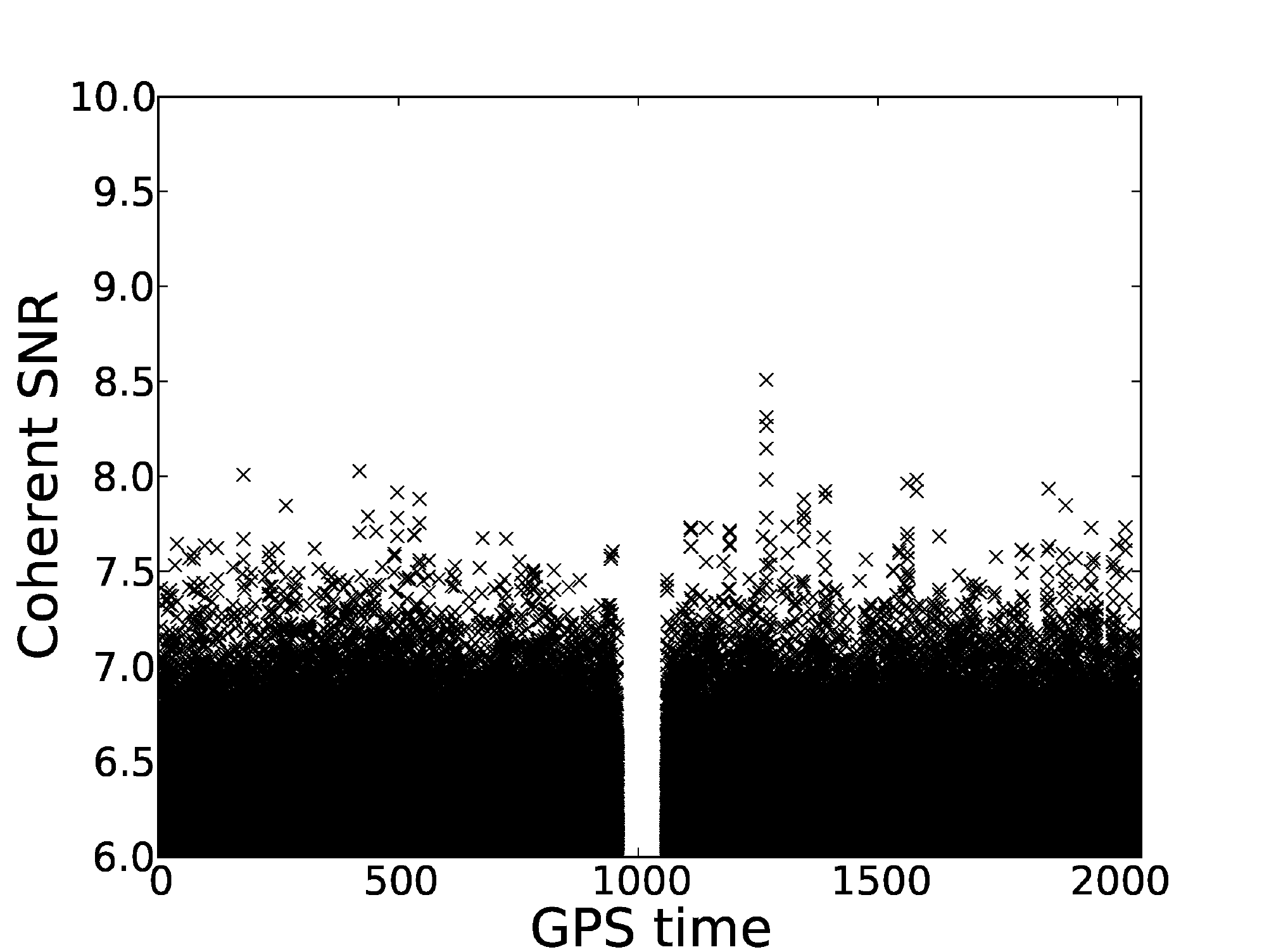}
  \end{minipage}
  \begin{minipage}[t]{0.495\linewidth}
    \includegraphics[width=\linewidth]{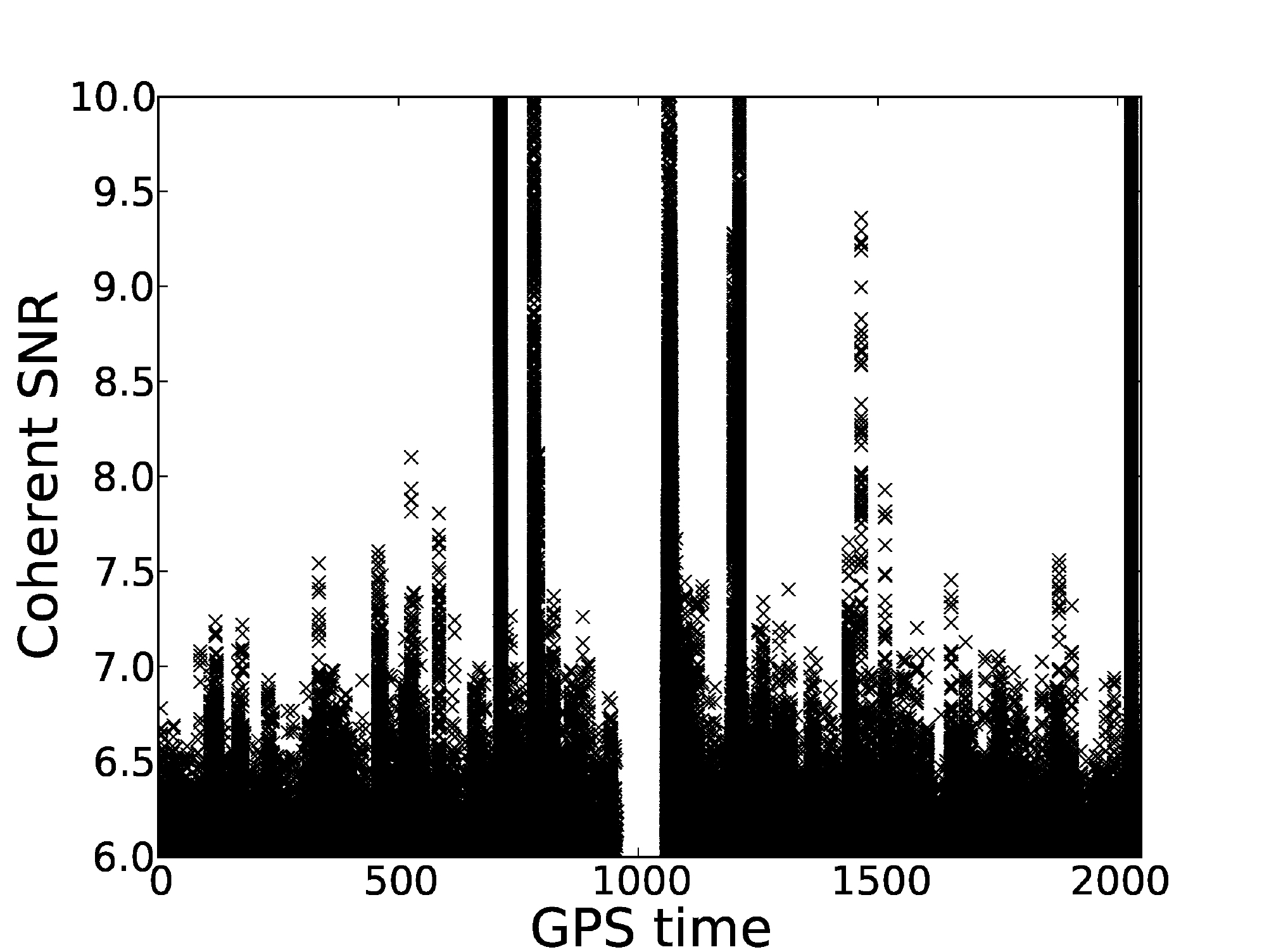}
  \end{minipage}
  \begin{minipage}[t]{0.495\linewidth}
    \includegraphics[width=\linewidth]{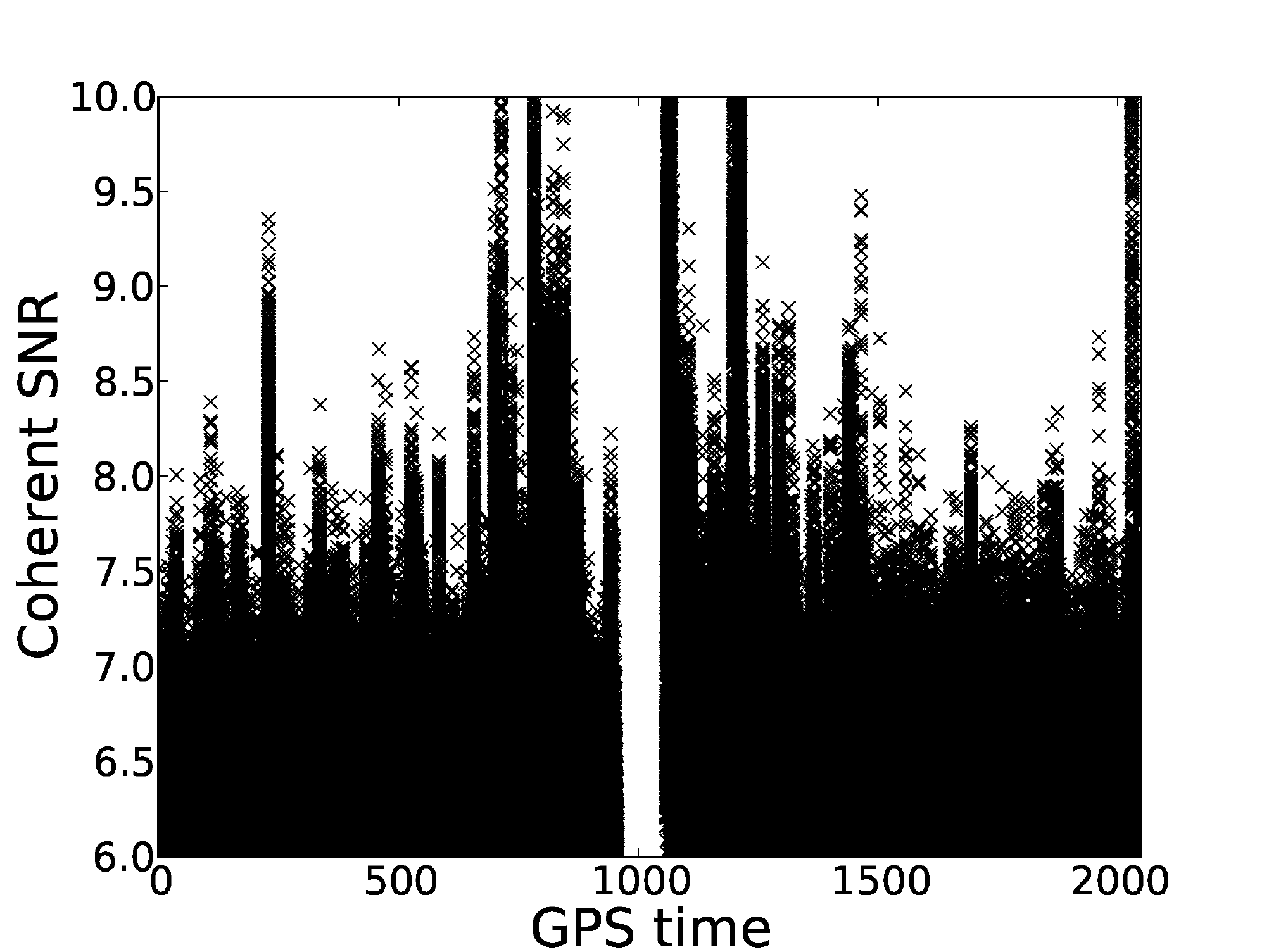}
  \end{minipage}
  \caption{\label{fig:S4results} 
The distribution of triggers found by the coherent \ac{PTF} search. The
left panels show the distribution of triggers from templates that were
analysed using the ``restricted'' coherent PTF search, the right panels
show the distribution from the templates that were analysed using the
``full'' coherent PTF search. The top panels were created from analysing
Gaussian noise.  The bottom panels were created from analysing a stretch
of real data from \ac{S4}. All these plots have been rescaled to use the
same y-axis. For the two cases using real data the non-Gaussian spikes
extend much higher than is shown, the loudest trigger has an \ac{SNR} of
39 in the restricted case and 45 in the full case.   }
\end{figure}

In Figure \ref{fig:S4results} we show the distribution of the \ac{SNR}
of the triggers produced using both the full and restricted statistic.
This is shown for the stretch of real \ac{S4} data and for a stretch of
simulated Gaussian data.  As expected, the \acp{SNR} of triggers in
Gaussian noise are larger for the precessing templates than the
restricted ones, even though significantly fewer templates were analysed
with the full statistic.  The results from real data are badly affected
by non-Gaussianities in the data.  A number of loud transients are
clearly visible as short duration peaks of large \ac{SNR}, while there
are an even greater number of quieter peaks throughout the analyzed
time.  This has a similar effect on both the full and restricted
waveforms.

In \cite{cohnonspin}, we described and developed a number of tools which
can be used to effectively remove the majority of the non-Gaussian
features from a non-spinning, coherent analysis.  These include null
stream consistency \cite{Guersel:1989th}, amplitude consistency and
$\chi^2$ signal consistency tests \cite{Allen:2004gu,Hanna:2008}.  All
of these can be applied without modification to the restricted \ac{PTF}
search, and it seems reasonable to expect they would be similarly
effective in reducing the effect of non-Gaussianities in the data.
However, we currently have no such tools which can be used for the full
\ac{PTF} waveforms.  Before using this search on real data we will need
to implement a set of tests that can discriminate glitches from real
signals for the full statistic.  It should be relatively straightforward
to implement the null stream consistency test.  Unfortunately,
as discussed in \cite{cohnonspin}, the null stream for the \ac{LIGO} S4
detectors is constructed only from the two instruments in Hanford. In
this stretch of data the loudest background triggers are caused by
non-stationarities in the Livingston detector and thus the null stream
is ineffective.  Alternatively a $\chi^2$ test such as the ones
described in \cite{Allen:2004gu,Hanna:2008} could be adapted to this
search, \cite{Fazi:2009} presents a possible way of doing this for
single detectors. We are working on developing an alternative version of
this $\chi^2$ test, which would test the consistency of the six
independent components of a single detector \ac{PTF} waveform, and then
extending this to the fully coherent analysis.

\section{Discussion}

In this paper we have presented a method for performing a coherent
search for precessing, single spin black hole--neutron star coalescences
using the \ac{PTF} method.  We have compared the performance to searches
using non-precessing waveforms and have identified regions of the
parameter space where the \ac{PTF} search offers increased sensitivity.
We have presented a method by which these areas could be identified and
demonstrated these techniques on a short stretch of S4 data.

This method should allow for the detection of highly precessing
\ac{NSBH} sytems with greater efficiency than the current non-spinning
searches. However, more work is required before this search is ready to
be used. The main need is for the development of effective methods of
separating glitches from real events in the full \ac{PTF} search,
whether performing a coincident or a coherent search.  As discussed in
section \ref{sec:search_and_results}, it should be possible to adapt a
lot of the methods that have proven effective in non-spinning searches
\cite{Allen:2005fk,cohnonspin} but this is a non-trivial task.

In this paper, we have focused on the \ac{PTF} precessing waveforms.
However, many of the techniques we have discussed would be equally
applicable to the dominant harmonic of \textit{any} family of precessing
waveforms.  In particular, the method of maximizing over freely over the
amplitudes of the five components of the $l=2$ spin weighted-spherical
harmonic is directly applicable to other waveform families.  As the
catalogue of numerical simulations of precessing binaries grows, these
methods may well find applications in searches using such as numerical
relativity inspired inspiral-merger-ringdown waveforms.  

\section*{Acknowledgements}

We would like to thank Duncan Brown, Sukanta Bose, Diego Fazi, Mark
Hannam, Drew Keppel, Andrew Lundgren, Ilya Mandel and Bangalore
Sathyaprakash for useful discussions and comments on this manuscript.
IWH was funded by the Science and Technology Facilities Council, UK,
studentship ST/F005954/1, SF would like to acknowledge the support of
the Royal Society.

\section*{References}

\bibliographystyle{unsrt}
\bibliography{references}

\end{document}